\documentclass[sigplan,screen,authorversion,nonacm]{acmart}

\usepackage{algorithm2e}
\usepackage{caption}
\usepackage{subcaption}
\usepackage{lipsum}
\usepackage{todonotes}

\AtBeginDocument{%
  \providecommand\BibTeX{{%
    \normalfont B\kern-0.5em{\scshape i\kern-0.25em b}\kern-0.8em\TeX}}}

\begin{document}

\title{Massively Distributed Finite-Volume Flux Computation}


\author{Ryuichi Sai}
\authornote{This author is a Ph.D. student at Rice University. This work was performed while the author was at TotalEnergies EP \mbox{Research \& Technology US, LLC.}}
\affiliation{%
  \institution{TotalEnergies EP Research \& Technology US, LLC.}
  \city{Houston}
  \state{Texas}
  \country{USA}
}
\email{ryuichi@rice.edu}

\author{Mathias Jacquelin}
\affiliation{%
  \institution{Cerebras Systems}
  \city{Sunnyvale}
  \state{California}
  \country{USA}
}

\author{Fran{\c c}ois P. Hamon}
\affiliation{%
  \institution{TotalEnergies EP Research \& Technology US, LLC.}
  \city{Houston}
  \state{Texas}
  \country{USA}
}

\author{Mauricio Araya-Polo}
\affiliation{%
  \institution{TotalEnergies EP Research \& Technology US, LLC.}
  \city{Houston}
  \state{Texas}
  \country{USA}
}

\author{Randolph R. Settgast}
\affiliation{%
  \institution{Lawrence Livermore National Laboratory}
  \city{Livermore}
  \state{California}
  \country{USA}
}



\renewcommand{\shortauthors}{Sai, et al.}

\begin{abstract}
Designing large-scale geological carbon capture and storage projects and ensuring safe long-term CO\textsubscript{2} containment -- as a climate change mitigation strategy -- requires fast and accurate numerical simulations.
These simulations involve solving complex PDEs governing subsurface fluid flow using implicit finite-volume schemes widely based on Two-Point Flux Approximation (TPFA).
This task is computationally and memory expensive, especially when performed on highly detailed geomodels.
In most current HPC architectures, memory hierarchy and data management mechanisms are insufficient to overcome the challenges of large scale numerical simulations.
Therefore, it is crucial to design algorithms that can exploit alternative and more balanced paradigms, such as dataflow and in-memory computing.
This work introduces an algorithm for TPFA computations that leverages effectively a dataflow architecture, such as Cerebras CS-2, which helps to significantly minimize memory bottlenecks.
Our implementation achieves two orders of magnitude speedup compared to multiple reference implementations running on NVIDIA A100 GPUs.

\end{abstract}



\keywords{finite-volume, 
two-point flux approximation, 
high-performance computing, 
energy, 
dataflow architecture, 
wafer-scale engine, 
distributed memory}



\maketitle

\section{Introduction}

Concentration of CO\textsubscript{2} in the atmosphere is reaching critical levels, reducing fossil fuel consumption or cleaning its production mechanisms is simply not enough, mitigation strategies are required to stay within United Nations targets.
Geological Carbon Capture and Storage (CCS) has been identified by International Panel for Climate Change (IPCC) as one of the ways
to reduce Greenhouse Gas (GHG) concentration.
Scaling up CCS capacity to match the targets requires to launch many new sites.
Designing these large-scale CCS projects require numerical simulations to ensure safety and long-term containment of the CO\textsubscript{2}.
Fast and accurate simulation is one of the key factors to design safe projects within regulatory and commercial time constraints.

The numerical simulation of physical processes involved in CCS requires solving complex PDEs governing fluid flow in the subsurface.
Commercial and research simulators are based on implicit Finite-Volume (FV) schemes widely using Two-Point Flux Approximation (TPFA) to discretize these equations.
These simulations are performed on highly detailed geomodels extremely computationally and memory demanding.
Unfortunately, in most current HPC systems, the imbalance between intricate memory hierarchies (host and devices) and networking results in poor overall computing performance~\cite{reed2022reinventing}, therefore limiting the performance of numerical simulations.
Consequently, it is critical to explore and leverage alternative architectures and corresponding algorithms allowing fast FV-based flow simulations.

HPC system design progress has led to algorithmic changes and optimizations in both scientific and business domains, with previous exploration of non-hierarchical architectures resulting in high computational efficiency~\cite{MauricioIBMCellBE,kahle1989connection}.
The raising of highly parallel systems with a distributed memory architecture that has significantly higher memory bandwidth, low memory latency, and lower energy cost for memory access thanks to dataflow-like architecture design (such as Cerebras, Groq \cite{groqHotChip}, SambaNova \cite{Samba21}, etc.) are emerging as alternatives to the traditional accelerated system.

In this work, we explore the capabilities of a dataflow architecture system with a novel implementation of the flux computation, which is fully based on localized communications and single-level memory, and we demonstrate breakthrough performance.
This paper makes the following contributions:

\begin{itemize}
\item it describes a novel TPFA implementation designed specifically for a dataflow architecture;
\item it introduces a data communication pattern on a single-level distributed memory architecture to support data movement from both cardinal and diagonal neighbors;
\item it describes various optimizations to speed up overall performance;
\item it presents both a dataflow implementation and reference GPU implementations based on different programming models; and
\item it evaluates performance on a Cerebras CS-2 machine and compares with the reference implementations on an accelerated system sporting NVIDIA A100 GPUs, and demonstrates strong performance characteristics.
\end{itemize}

This paper follows the organization outlined below.
The paper begins by reviewing related work in Section~\ref{sec:related}.
Section~\ref{sec:geosx} outlines the problem domain.
Section~\ref{sec:architecture} provides a discussion about dataflow architectures, while 
in Section~\ref{sec:tpfacs2}, we describe our approach to map the flux calculation algorithm to a dataflow architecture.
Section~\ref{sec:tpfaraja} presents our reference implementations for GPUs.
We evaluate and compare these implementations in Section~\ref{sec:evaluation}.
Finally, we provide discussions and conclusions in Sections~\ref{sec:discussions}~and~\ref{sec:conclusion}.

\section{Related Work}\label{sec:related}

\paragraph{Finite-Volume for geologic CCS simulation}

Implicit finite-volume schemes \cite{lie2019introduction} are widely used to simulate flow and transport in geologic porous media. 
The computation of the intercell flux and its derivatives with respect to the state variables is a key step of the simulator workflow.
In the present work, we exploit a dataflow architecture to accelerate a simplified finite-volume flux calculation kernel.
This is a preliminary step towards the porting of the full set of discretized flow equations to the dataflow model.

These efforts could ultimately benefit to reference CCS simulators like GEOS \cite{GEOSXWebsite}.
GEOS is an open-source, multiphysics C++ computational platform for the simulation of CCS and other subsurface energy systems \cite{cusini2022field,camargo2022deformation,huang2022validation,costa2022multi,bui2021multigrid}.
%
GEOS uses a coupled finite element – finite volume formulation to simulate thermal multiphase flow and mechanics with faults and fractures.
Its design emphasizes modularity and extensibility, making it easy to add new solvers or modify existing ones.
The code leverages the RAJA library for portable execution policies and targets massively parallel CPU and GPU architectures.
%



\paragraph{Accelerating applications on dataflow architecture}

The utilization of dataflow architecture and its implementations for HPC workloads was initially explored in a previous study~\cite{DirkSC20}, which achieved 0.86 PFLOPS on a Cerebras CS-1 system for a 7-point stencil on a $600\times595\times1536$ mesh.
While a refreshed version~\cite{woo2022disruptive} of~\cite{DirkSC20} is evaluated on CS-2, it mainly focuses on a Python interface to address programming productivity issues related to low-level abstractions.
Our study is inspired by this work, but differs in our use of Cerebras Software Language (CSL) for implementation, which offers a higher level of abstraction and enables reusable functions and modules.
Additionally, our problem domain involves more complex computations and data communication patterns. Further, our study focuses on achieving maximum performance on a real-world application that involves intricate data communication and requires computation with higher arithmetic intensity.

In~\cite{MauricioSC22}, a memory layout is proposed for efficient computing of a large 3D data grid on a dataflow architecture, taking full advantages of its on-chip memory and 2D fabric.
Furthermore, a communication strategy is designed to exploit the architecture's fast communication fabric.
Computing the physics in our work demands more memory footprint and involves intricate data communication than that in~\cite{MauricioSC22}.
Additionally, our problem extends data movement to diagonal neighbors in addition to adjacent cardinal neighbors, which has not been explored, and leverages several optimizations to accelerate the performance.

Other dataflow-like architectures have been introduced, in particular, 
in~\cite{Samba21b} SambaNova presents a set of relevant results for scientific applications from the Machine Learning perspective. Another entry~\cite{louw2021using} of the dataflow-oriented approaches for HPC scientific computing is Graphcore, which explores high operating intensity applications, such as stencil-based Lattice-Boltzmann and Gaussian filter.



\section{Finite Volume for Geologic Carbon Storage Simulation}\label{sec:geosx}

In this work, we consider a compressible single-phase flow Darcy-scale model able to represent the pressure changes due to fluid (supercritical CO$_2$) injection in a geological formation.
The system of governing equations consists of Darcy's law and a mass balance equation:
\begin{subequations}
\begin{align}
& \boldsymbol{u} = - \frac{\kappa}{\mu} ( \nabla p - \rho \boldsymbol{g} ), & & \mbox{(Darcy's law),} \label{eq:single_phase_darcy} \\
& \frac{\partial}{\partial t} \left( \phi \rho \right) + \nabla \cdot \left( \rho \boldsymbol{u} \right) = 0, & & \mbox{(mass balance).} \label{eq:single_phase_mass_balance}
\end{align} 
\end{subequations}
where $\boldsymbol{u}$ is the Darcy velocity, $\kappa$ is the (scalar) permeability coefficient, $p$ is the pressure, and $\boldsymbol{g}$ is the gravity vector weighted by gravitational acceleration.
The porosity, $\phi$ and the density, $\rho$, depend linearly on pressure.
The viscosity, $\mu$, is assumed to be constant.

It is common to discretize this near-elliptic system by combining a low-order FV scheme with an implicit (backward-Euler) temporal discretization.
Considering a mesh consisting of $n$ cells, the discretized system of nonlinear equations is:
\begin{equation}
V_K \frac{ \phi^{n+1}_K \rho^{n+1}_K - \phi^{n}_K \rho^{n}_K }{\Delta t} + \sum_{L \in \text{adj}(K)} F_{KL}^{n+1} = 0,
\label{eq:vk}
\end{equation}
where $V_K$ is the volume of cell $K$, $\phi^{n+1}_K$ and $p^{n+1}_K$ are respectively the porosity and the pressure in cell $K$ at time $n+1$, and $\Delta t$ is the time step size.
In this work, we focus on the calculation of the discrete interfacial flux, denoted by $F_{KL}^{n+1}$, and neglect the accumulation term.
We use a standard TPFA method combined with single-point upwinding, in which case the flux reads:
\begin{subequations}
\begin{align}
F^{n+1}_{KL} &= \Upsilon_{KL} \lambda_{\text{upw}}^{n+1} \Delta \Phi^{n+1}_{KL}, \\
\Delta \Phi^{n+1}_{KL} &=  p^{n+1}_L - p^{n+1}_K + \rho^{n+1}_{\text{avg}} g ( z_L - z_K ),
\end{align} \label{eq:flux-calc}
\end{subequations}
where $\rho^{n+1}_{\text{avg}}$ is the average of the fluid densities in cells $K$ and $L$ at time $n+1$ and $z_L - z_K$ is the elevation difference between the centers of cells $L$ and $K$.
The transmissibility, $\Upsilon_{KL}$, is a coefficient accounting for the geometry of the cells and their permeability.
The upwinded fluid mobility, $\lambda_{\text{upw}}^{n+1}$, is computed as a function of the density, viscosity, and potential difference as:
\begin{equation}
\lambda_{\text{upw}}^{n+1} =
\begin{cases}
\displaystyle \frac{\rho^{n+1}_K}{\mu} \quad \text{if} \, \, \Delta \Phi^{n+1}_{KL} > 0 \\
\displaystyle \frac{\rho^{n+1}_L}{\mu} \quad \text{otherwise}
\end{cases}
\end{equation}
The fluid density is evaluated using the cell pressure using a slight compressibility assumption:
\begin{equation}
    \rho^{n+1}_K = \rho_{\text{ref}} e^{c_f ( p^{n+1}_K - p_{\text{ref}})},
    \label{density}
\end{equation}
where $c_f$ denotes the fluid compressibility, $p_{\text{ref}}$ is the constant reference pressure, and $\rho_{\text{ref}}$ is the constant reference density.

\begin{algorithm}[ht]
\KwResult{Flux part of residual, $\mathbf{r}_{\text{flux}}$}
$\mathbf{r}_{\text{flux}} := \mathbf{0}$\;
\For{each cell $K$}{
    \For{each neighbor cell $L \in \text{adj}(K)$}{
        Evaluate densities in $K$ and $L$ using Eq.~\ref{density}\;
        Evaluate flux $F_{KL}$ using Eq.~\ref{eq:flux-calc}\;
        Increment the residual in cell $K$: $(\mathbf{r}_{\text{flux}})_K = (\mathbf{r}_{\text{flux}})_K + F_{KL}$;
      }
    }
    \caption{FV flux computation algorithm considered in this work.}
    \label{algo:tpfa}
\end{algorithm}

The outer-loop of Algorithm~\ref{algo:tpfa} sweeps over cells and its inner-loop traverses over the neighbors of each cell.
For each neighbor, a flux is computed and is used to increment the local residual in the cell.
Here, we consider a 3D implementation of the algorithm on a Cartesian mesh. 
Each interior cell therefore has six neighbors (four neighbors in the same X-Y plane as the cell, one neighbor at the top and one at the bottom) and involves the computation of six fluxes. In our implementation, we also compute four fluxes between a cell and its diagonal neighbors to prepare the communication pattern for either higher-accuracy schemes or more intricate meshes. 
Algorithm~\ref{algo:tpfa} can be applied to unstructured meshes but will require a more sophisticated communication pattern to do so.
Algorithm~\ref{algo:tpfa} is applied 1,000 times with a different pressure vector at every call.


\section{An Alternative Approach to Computing}
\label{sec:architecture}

\begin{figure}[ht]
  \centering
  \includegraphics[width=.85\linewidth]{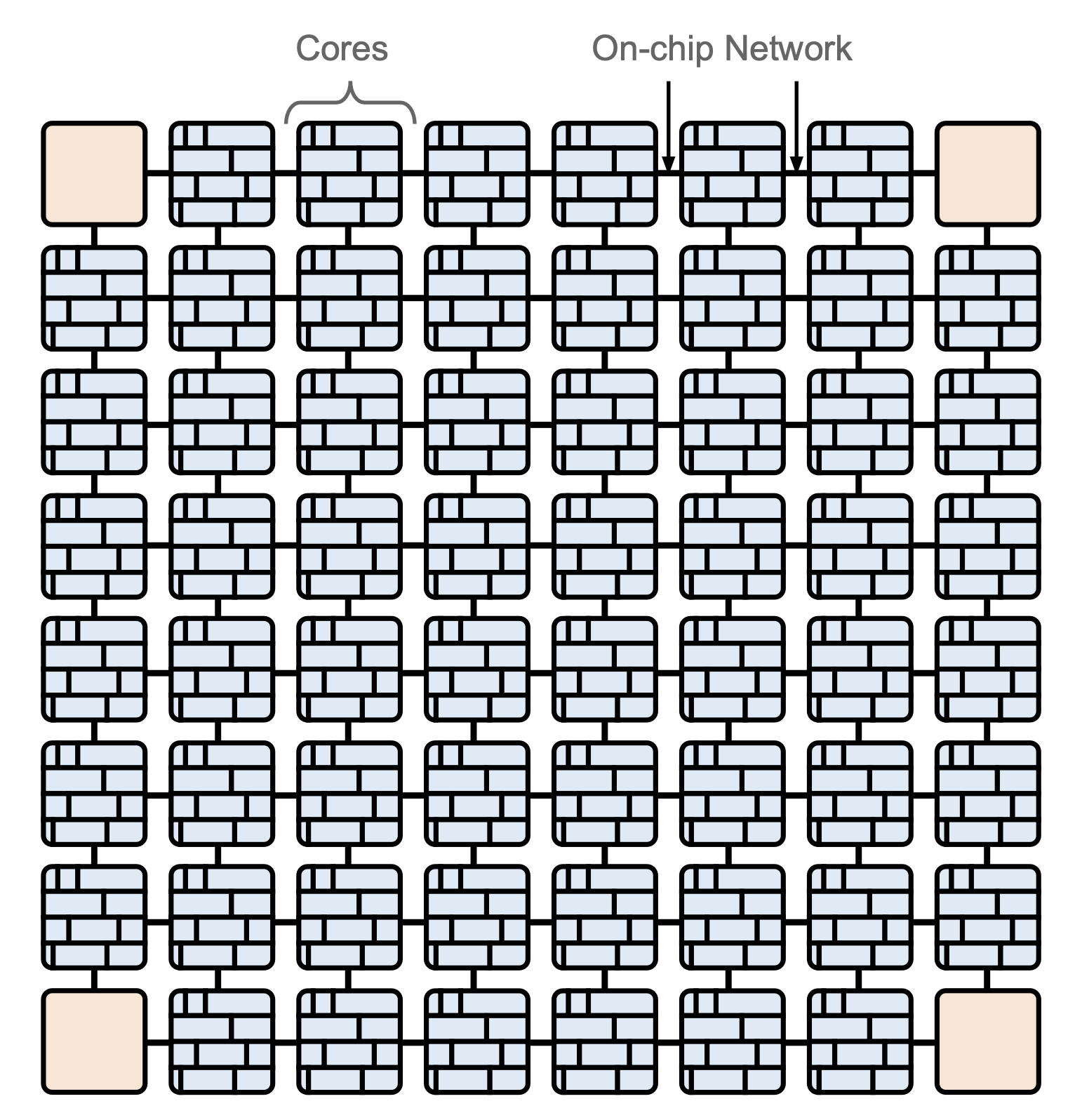}
  \caption{A typical implementation of a dataflow architecture. Figure from \cite{groqHotChip}.}\label{figure:groq}
\end{figure}

In this work, one of the main contributions is the effective use of an architecture that combines elements of the dataflow computing and the in-memory computing paradigms, plus access to a large amount of computing elements concurrently.
Figure~\ref{figure:groq} shows a typical implementation for such architecture, and 
in \cite{groq22b} some of these topics are explored. For the remainder of this work and for brevity, we will refer to this kind of emerging architectures as dataflow architectures, but in later sections one specific architecture will be discussed prior to introducing the implementation details.
The main disruptive aspects are the independence from deep memory hierarchies, and the locality of the available memory with respect to computing units.
For instance, all compute and memory resources are embedded on the same wafer and are interconnected via a low latency/high throughput fabric. 

The mapping of algorithms on such architecture needs to consider two main aspects, in hierarchical fashion: first, a top-level hierarchy concern is to effectively distribute data among a large number of processing units with non-centralized coherent memory interface. Second, once data is in the local memory of the computing units, the lower-level challenge is to use it as much as possible before sharing with neighboring computing units. The top-level concern is to efficiently keep data flowing between processing elements, on a more traditional architecture, this is the level that would be usually implemented with MPI. The lower-level concern is utilizing a scratch-pad like memory system efficiently, as opposed to traditional architectures, where the main challenge -- for a memory bandwidth-bounded algorithm -- is to exploit the complex memory hierarchies. Compounded to that fact, most current HPC systems force developers to be aware of a full memory hierarchy on both the host and the device sides of accelerated computing nodes. In traditional architectures, the latter is usually addressed with CUDA/HIP, or highly portable programming models like RAJA/Kokkos or OpenMP.

\section{Finite-Volume Flux Computation on a Dataflow Architecture}\label{sec:tpfacs2}

The section details the implementation of a FV flux computation using a TPFA on a dataflow architecture.
First, we outline the mapping of both data and algorithm to the fabric.
Next, we describe the data movement patterns (including both between cardinal neighbors and diagonal ones) and TPFA computation for the horizontal plane.
Then, the algorithm is scaled on the third dimension to support numerous planes.
Subsequently, the implementation is expanded for multiple iterations.
Lastly, we elaborate on the optimization techniques used to speed up performance.

\begin{figure}[ht]
  \centering
  \includegraphics[width=.9\linewidth]{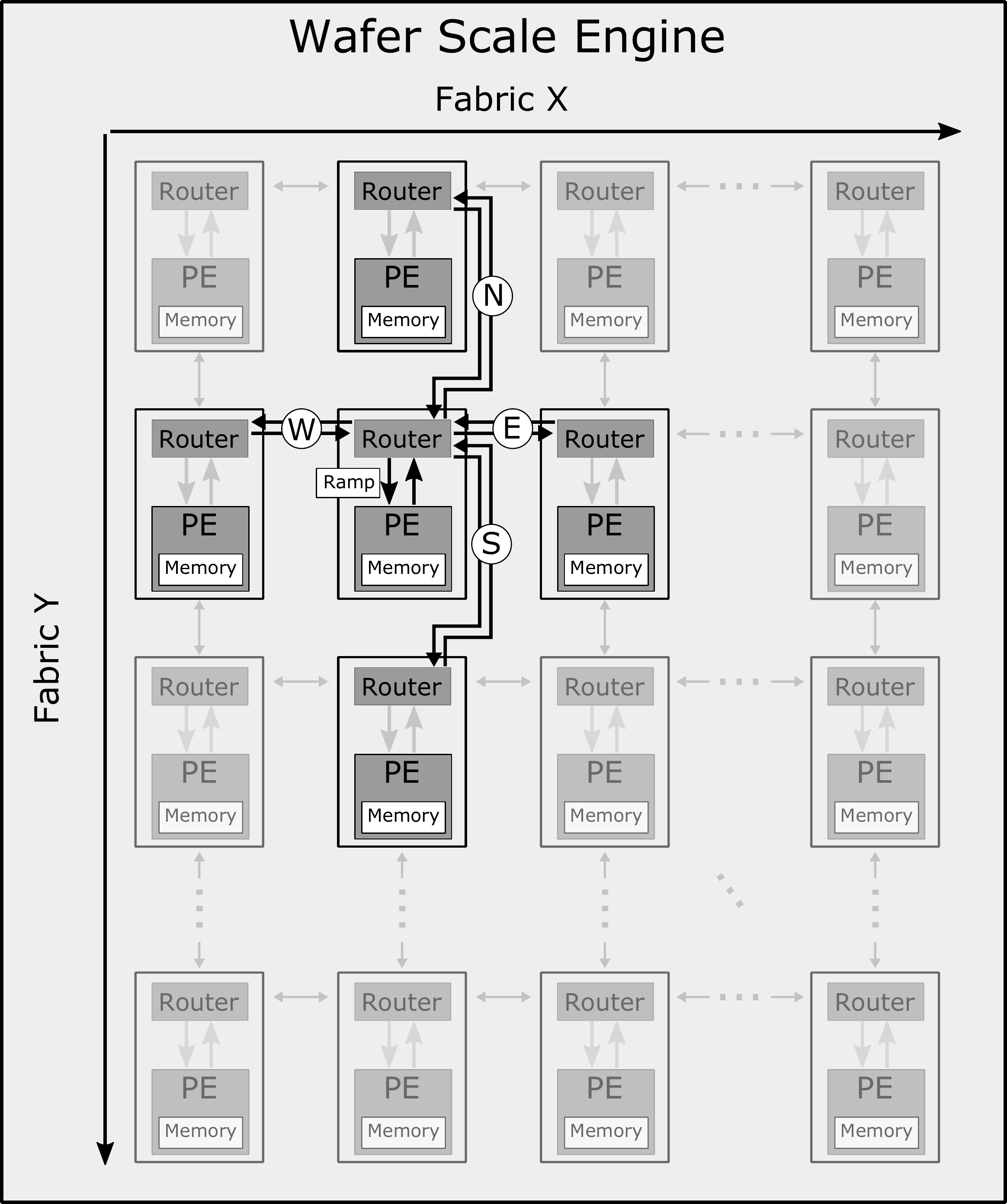}
  \caption{A simplified overview of the Cerebras Wafer Scale Engine}\label{figure:wse}
\end{figure}

Figure~\ref{figure:wse} provides a simplified overview of the Cerebras Wafer Scale Engine (more details are available in~\cite{MauricioSC22}).
The Wafer Scale Engine (WSE) is an example of a dataflow architecture, which as most dataflow architecture implementations, comes with a 2D-mesh interconnection fabric that connects processing elements (PEs) where computations take place.
Each PE has its own private local memory and is connected to a router.
The router manages five full duplex links: 
1) North, East, South, and West links connect to neighboring routers, allowing data to flow between PEs;
2) The Ramp link allows data to flow between a PE and its own router.
Each of these links transfers data in 32-bit packets. 
Each packet is associated with a color, or tag, used for routing and indicating the type of a message.

\subsection{Data Mapping}

In this study, the physical problem is represented by a 3D Cartesian mesh, where each cell has flux connections with 10 neighbors.
On the X-Y plane (horizontal), each cell has eight flux connections, four of which align with the cardinal axes and the other four with the diagonals.
For each flux connection, a local flux is computed and assembled with all other fluxes of the same cell.

\begin{figure}[ht]
  \centering
  \includegraphics[width=.6\linewidth]{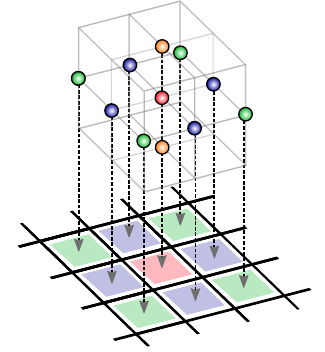}
  \caption{Three dimensional problem mapping to two dimensional fabric of processing elements using a cell-based approach.}\label{figure:cell-based-mapping}
\end{figure}

To compute the flux for all the cells in the mesh, two mapping techniques are considered, namely, face-based and cell-based mappings, as shown in Figure~\ref{figure:cell-based-mapping}.
The latter approach is the most straightforward to map to fabric with each cell mapping to a PE. 
This allows to leverage the architecture's computation, memory, and communication resources.

\begin{figure}[ht]
  \centering
  \includegraphics[width=.6\linewidth]{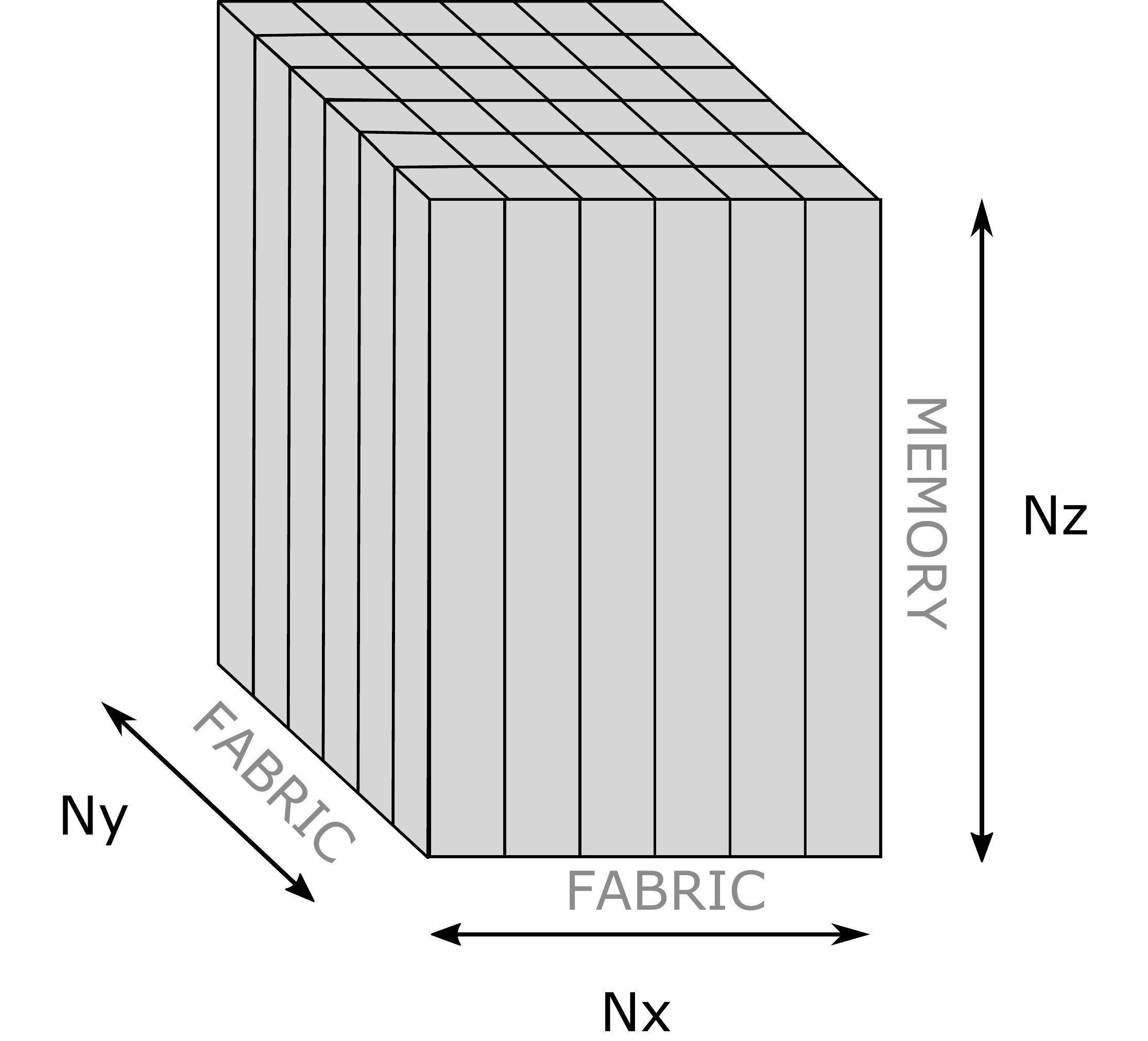}
  \caption{3D grid of size ${N_x \times N_y \times N_z}$. X and Y dimensions are mapped onto the PE grid of the WSE, while Z dimension is mapped onto memory of each PE.}\label{figure:pe-mapping}
\end{figure}

\begin{figure*}[t]
  \centering
  \begin{subfigure}[t]{0.3\textwidth}
      \centering
      \includegraphics[width=\textwidth]{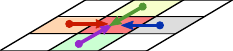}
      \caption{Step 1.a - North, South, East, and West PEs send to the target PE.}
      \label{fig:comm_phase1}
  \end{subfigure}
  \hfill
  \begin{subfigure}[t]{0.3\textwidth}
      \centering
      \includegraphics[width=\textwidth]{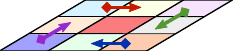}
      \caption{Step 1.b - Corner PEs send to North, South, East, and West PEs (concurrent to Step 1.a).}
      \label{fig:comm_phase2}
  \end{subfigure}
  \hfill
  \begin{subfigure}[t]{0.3\textwidth}
      \centering
      \includegraphics[width=\textwidth]{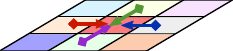}
      \caption{Step 2 - North, South, East, and West PEs forward data from Corner PEs to target PE.}
      \label{fig:comm_phase3}
  \end{subfigure}
     \caption{Communication pattern in the X-Y plane. The target PE (in red) receives data from all 8 surrounding PEs.}
     \label{fig:pe-common-patt}
\end{figure*}

The data domain is a 3D Cartesian mesh with dimensions of ${N_x \times N_y \times N_z}$, depicted in Figure~\ref{figure:pe-mapping}.
We decompose the data domain such that every cell from the Z-dimension is mapped to the same PE, while the X and Y dimensions are mapped across the two axes of the fabric.
To be more precise, we map a cell with coordinates $(x, y, z)$ in the 3D mesh onto PE $(x, y)$.
The whole Z-dimension resides in the local memory of a PE. 
This strategy, as previously demonstrated in~\cite{DirkSC20,MauricioSC22}, optimizes the utilization of dataflow architecture and maximizes the potential parallelism for HPC applications of this kind.

Each flux computation requires three types of data: the cell-centered data of the current cell, the cell-centered data of a neighboring cell, and the face-centered data for each pair of cells.
Figure~\ref{figure:cell-based-mapping} also depicts cell dependencies.
The center cell, along with its top and bottom neighbors are mapped onto the PE colored in red.
Its cardinal neighboring cells and corresponding PEs are marked in purple while diagonal cells and PEs are depicted in green.
Each PE allocates memory space for its current residual, pressure, and gravity coefficients, as well as 10 transmissibilities for the fluxes between the cell and its neighbors.
Each PE also allocates space to receive the pressure and gravity coefficients from all eight neighboring cells on the same X-Y plane at each iteration.
More details on the communication pattern will be given below.


\subsection{Distributed Data Communication in Flux Computation}

As mentioned earlier, computing a specific cell depends on data from 10 neighboring cells in all three dimensions.
Precisely, a cell $(x, y, z)$ requires:

(a) cells from X-Y cardinal neighbors, corresponding to cells $(x+1,y,z)$,
 $(x-1,y,z)$,
 $(x,y+1,z)$, and
 $(x,y-1,z)$;

(b) cells $(x+1,y+1,z)$,
 $(x-1,y-1,z)$,
 $(x-1,y+1,z)$, and
 $(x+1,y-1,z)$ from X-Y diagonal neighbors; and

(c) cells $(x,y,z+1)$ and
 $(x,y,z-1)$ along the Z-dimension.


\subsubsection{Data communication from X-Y cardinal neighbors}\label{subsection:data-move-cardinal} \hfill\\

In the X-Y plane, a PE has to communicate with its immediate neighbors along each cardinal direction of the fabric.
Specifically, for the X-dimension, a PE must communicate with its eastbound neighbor at cell $(x+1,y,z)$ and its westbound neighbor at cell $(x-1,y,z)$.
For the Y-dimension, a PE must communicate with its northbound neighbor at cell $(x,y-1,z)$ and its southbound neighbor at cell $(x,y+1,z)$.
Similar to the communication strategy adopted in~\cite{MauricioSC22},
Two dedicated colors are assigned for each communication pattern, one for sending data and another for receiving data.
Upon receiving the data, the corresponding flux computation will occur immediately in an asynchronous fashion. 
This strategy allows overlapping communications with computations.


In each communication phase, multiple PEs concurrently send data out to all four directions, with each \textit{Sending} PE sending its local block of data of length ${N_z \times 2}$ on each respective direction.
Figure~\ref{fig:comm_phase1} shows that each PE concurrently receives data from all four directions.
Each \textit{Receiving} PE receiving a total length of ${N_z \times 2 \times 4}$ from four neighbors.

\begin{figure*}[t]
  \centering
  \begin{subfigure}[b]{0.3\textwidth}
      \centering
      \includegraphics[width=.35\textwidth]{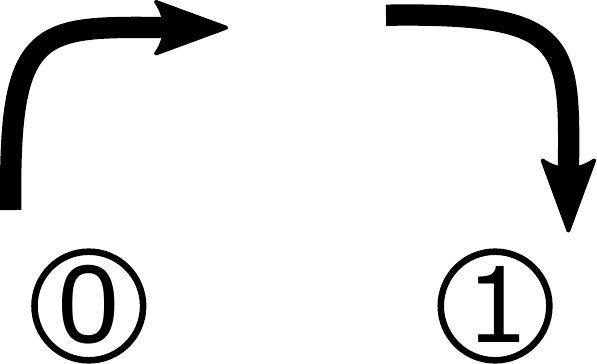}
      \caption{Router configurations used by FV flux computation. Configuration 0 corresponds to the configurations of a \textit{Sending} PE as the root of a broadcast, configuration 1 is the used by a \textit{Receiving} PE.}
      \label{fig:pe-router-configurations}
  \end{subfigure}
  \hfill
  \begin{subfigure}[b]{0.6\textwidth}
      \centering
      \includegraphics[width=.6\textwidth]{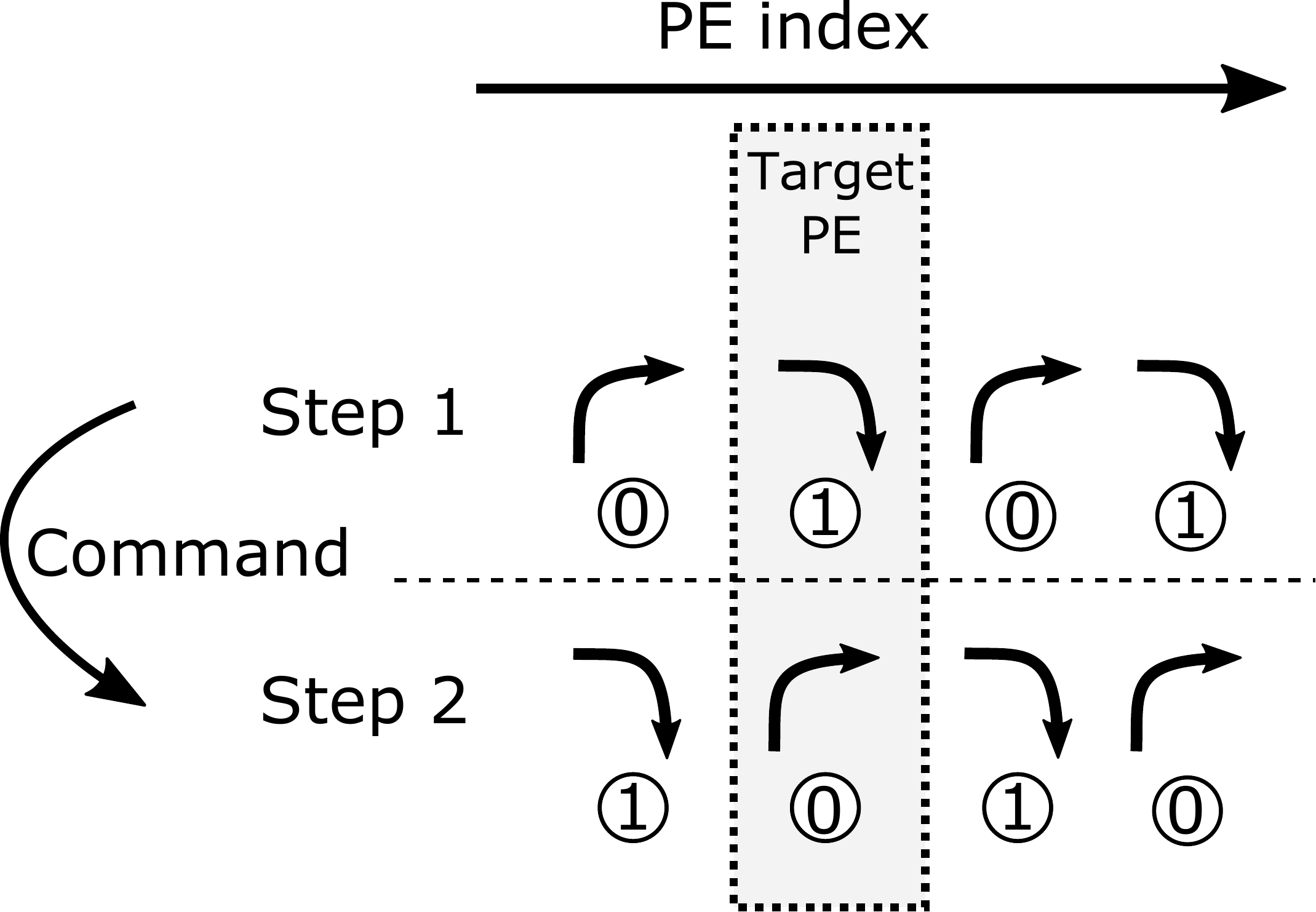}
      \caption{Two communication steps required to fetch all the data required by a target PE from the West. Corresponding router configurations are given in the circled numbers. At each step, a router command is sent through the broadcast pattern, changing the configurations from one to the alternative router configuration.}
      \label{fig:pe-communication-steps}
  \end{subfigure}
     \caption{Eastward localized broadcast operation used in FV flux computation to exchange cell values along the X dimension.}
     \label{fig:pe-eastward-broadcast}
\end{figure*}

To switch a PE from \textit{Sending} to \textit{Receiving}, we control the PE router and change its configurations at runtime.
As shown in Figure~\ref{fig:pe-router-configurations}, two switch positions are defined for each PE for sending and receiving accordingly.
After sending its local data, a \textit{Sending} PE issues a command to switch the configurations of itself and the neighboring routers, such that
the original \textit{Sending} PE then becomes the \textit{Receiving} PE, and the original \textit{Receiving} PE becomes the \textit{Sending} PE.
Afterwards, the new \textit{Sending} PE then sends its local data and switching command to itself and its neighbors, and for the continuation of the experiment.
Since the data and command are sent and received from the immediate neighbors, the PE can alternate between a \textit{Sending} and a \textit{Receiving} state, as shown in Figure~\ref{fig:pe-communication-steps}.
After two steps, all data have been sent and received by all PEs.

\subsubsection{Data movement from X-Y diagonal neighbors}\label{subsection:data-move-diagonal} \hfill\\


To the best of our knowledge, we are the first to explore data movement from diagonals on a dataflow architecture.
Although this is not mandatory for evaluating the mathematical scheme presented in Section~\ref{sec:geosx}, we implemented it because we deem necessary to prepare for more intricate communication patterns.
Data communication along diagonals presents its own challenges due to the fabric layout being a perfect 2D grid.
The routing capabilities per PE are constrain to cardinal directions.
Therefore, while data communication along direct cardinal neighbors only involves two adjacent PEs, data transfer among diagonals requires multiple PEs with some serving as intermediaries.
An intermediary cell must be an immediate neighbor to both the source cell and its diagonal destination cell, and must share a cardinal axis with respect to these two cells.


As shown in Figure~\ref{fig:comm_phase2}, to transfer data from the northwest corner cell to the center cell, we first transfer the data from the northwest cell to the intermediary north cell, using an eastbound channel. We then transfer the data from north cell to the center cell, using a southbound channel, as shown in Figure~\ref{fig:comm_phase3}.


To enable concurrent communication from all four diagonal neighboring cells, 
a rotating and coordinating synchronization mechanism is used. 
This mechanism involves each corner PE sending data either clockwise or counterclockwise. 
As long as all four cells follow the same direction and each is assigned to a different intermediary cell, data transfers from all corner PEs are performed concurrently.



From an implementation perspective, data communication from X-Y cardinal neighbors remained unchanged, where each \textit{Sending} PE simply sends data to all four directions.
For an \textit{Intermediary} PE, it performs the partial flux computation for the face with its adjacent neighbor cell while sending the received data to the \textit{Receiving} PE.
The \textit{Receiving} PE, as illustrated in Figure~\ref{fig:pe-common-patt}, receives data from four cardinal directions twice:
initially, it performs a partial flux computations for these four faces with its direct cardinal neighbors;
and subsequently, the data from its diagonal neighbors arrives, and it performs the rest of the flux computations for the four faces with its diagonal neighbors.

\subsection{Optimizations}


We highlight some of the most effective and impactful optimizations in terms of computing performance and memory utilization.
Let's briefly remind that the cells in the same vertical column share the private memory of a PE, therefore reducing the memory consumption on each PE is crucial to fit the largest possible problem. 
With respect to performance, asynchronous communications between PEs are necessary to overlap data movement with computations. Further, vectorization is used to exploit the bandwidth of the fabric. 

\subsubsection{PE private memory saving strategies} \hfill\\


In local private memory, PE needs to store the instructions.
In addition, to storing cell and face data, the local memory must also accommodate extra buffers for data broadcasting and computations. 
In this implementation, data buffers reuse is critical,
this is akin to register allocation optimization, which is usually performed by compilers, but hand crafted in this work.
The impact is twofold: by minimizing the amount of memory the implementation requires, larger problems can be solved. Additionally, in some situations, overwriting / reusing data buffers eliminates the need for data replication, leading to more efficient computations as well.




\subsubsection{Asynchronous communications} \hfill\\
The implementation of the flux calculation algorithm relies on asynchronous communications to exchange data between processing elements. The data is sent/received asynchronously along four directions. 
This has two main benefits. First, simultaneous transfers increase the chance to exploit interconnect-level concurrency. Second, non-blocking communications allows to overlap transfers with useful computations, effectively hiding associated overheads as much as possible since the fabric and routers work completely independently from the processing elements.



\subsubsection{Vectorization of floating-point operations} \hfill\\


Most hardware architecture offer dedicated mechanisms to process arrays of data (or vectors). 
In the architecture at hand, this is implemented by using special registers holding (\textit{Data Structure Descriptors (DSD)}), that act as vectors, on which a given instruction can operate upon. The DSD contains information about the address, length, and stride of the arrays on which the instruction will operate. This allows a vectorized instruction to fully utilize SIMD units (up to 2 in single precision arithmetic) at hardware level. In a way, using data flow terminology, an instruction operating on DSDs acts as a filter through which data is flowing. Remarkably, no matter how long the input and output arrays are, the throughput of the instruction will be constant since there is no cache that can be filled up.
Consequently, the impact of using these dedicated registers significantly enhanced performance. 



\section{Finite-Volume Flux Computation on a GPU}\label{sec:tpfaraja}

This section describes our reference FV flux implementation on a GPU as a baseline when comparing to the one on a dataflow architecture.

The reference implementation employs a standard implicit FV flux calculation based on TPFA.
The numerical method is similar to those adopted by commercial simulators and benchmark open-source simulators such as GEOS~\cite{GEOSXWebsite}.
To fairly compare to our implementation on a dataflow architecture, our reference implementation uses a cell-based looping pattern.
The reference implementation with the cell-based approach is optimized to make it as fast as possible on a GPU.



To begin, we allocate spaces on both host memory and device memory.
We then load our data mesh of size $N_x \times N_y \times N_z$  into host memory, with the X-dimension as the innermost dimension and Z-dimension as the outermost dimension in the memory layout.
Next, we copy all data from host memory to device memory.
Since we evaluate our GPU kernel on the latest hardware with large enough device memory to load all data at once,
we avoid data domain decomposition and save time from frequent data transfer between host memory and device memory.

For each application of Algorithm~\ref{algo:tpfa}, we launch a GPU kernel with 3D threadblocks, carefully mapping each cell to a GPU thread.

\begin{figure}[t]
  \centering
  \begin{verbatim}
RAJA::KernelPolicy<
 RAJA::statement::CudaKernelFixed< 16 * 8 * 8,
  RAJA::statement::Tile<1, RAJA::tile_fixed<8>, 
                        RAJA::cuda_block_y_direct,
  RAJA::statement::Tile<0, RAJA::tile_fixed<16>, 
                        RAJA::cuda_block_x_direct,
   RAJA::statement::For<2, RAJA::cuda_block_z_direct,      
   RAJA::statement::For<1, RAJA::cuda_thread_y_direct,   
   RAJA::statement::For<0, RAJA::cuda_thread_x_direct, 
    RAJA::statement::Lambda<0>>>>>>>>;
\end{verbatim}
  \caption{RAJA execution policy used in kernel launch.}
  \label{figure:raja-policy}
\end{figure}

Our kernel launch is RAJA~\cite{RAJAPaper}-based,
which is widely used in physical simulations for maximum portability.
The looping pattern is implemented using a RAJA kernel policy, as shown in Figure~\ref{figure:raja-policy}, on a nested loop space of the whole data mesh size of $N_x \times N_y \times N_z$.
We instruct RAJA to launch kernels with GPU threadblock size of $1024$ to respect the GPU's limit of at most $1024$ threads per block, while maximizing the thread parallelism and GPU utilization.
We also guide RAJA to tile the GPU threadblock to $16\times8\times8$, where $16$ is the innermost dimension size.
Finally, we apply RAJA's GPU thread looping policies using \textit{cuda\_thread\_z\_loop},
\textit{cuda\_thread\_y\_loop},
\textit{cuda\_thread\_x\_loop} policies on three respective dimensions.

RAJA provides a C++ lambda with each index $(x, y, z)$ directly mapping to a cell in our data mesh.
Each GPU blockthread is scheduled to concurrently invoke a device function that performs the FV flux computation for its respective mapping cell.
First, each thread concurrently fetches the cell data for itself and all cell data from its ten neighboring cells.
Next, for each neighbor, it performs a flux computation using the transmissibility, the local cell values, and its neighbors values, and produces a local flux value.
Then, it assembles all the local fluxes and updates the current cell value.

The functions that perform the flux computation in the reference implementation are logically identical to those in our dataflow implementation.
As both implementations are written in C-like languages, the flux function code also look syntactically similar.
The only difference is that, unlike the memory space for each PE is isolated in dataflow architecture, GPU device memory is shared among all threads.
Therefore, we do not need to transfer the data among cells and can directly refer to the data using simple index arithmetic.
After the kernel completes all the applications of Algorithm~\ref{algo:tpfa}, data is copied from device memory back to host memory.

To validate the numerical accuracy and performance validity, we developed a second GPU kernel using the CUDA programming model manually.
The hand-crafted CUDA version has the same memory layout, uses the same tile sizes, and performs the same FV flux computation.
However, it launches its kernels with manually calculated block dimension and calculates the index mapping to the cell carefully.
It also needs to handle boundary checking to ensure the cell is still within the data grid.

\section{Experimental Evaluation}\label{sec:evaluation}

We evaluate our dataflow implementation on a Cerebras \hbox{CS-2} system and compare it to two reference implementations running on a NVIDIA A100 GPU. We discuss their respective performance characteristics.

\subsection{Experimental Configurations}


The CS-2 is the second generation chassis developed by Cerebras, which incorporates the 7nm technology and equips with total of 850,000 processing elements on a WSE-2 fabric.
Our experiments use maximum of the fabric of size $750 \times 994$ to utilize the massively distributed computing power.
This is the maximum size allowed by the SDK as a think layer of PEs around the boundary of the fabric are reserved for its own need.
While the CS-2 is connected to a Linux server, it is only used to schedule the workload, and no computations take place on the Linux machine during the experiments. 
We used Cerebras SDK 0.6.0, which includes a faster compiler, additional utility libraries, such as \textit{memcpy}, 
and offers improved performance compared to the previous version reported in~\cite{MauricioSC22}.

The GPU-based platform used as a reference is \textit{Cypress} from TotalEnergies.
The host machine has a 16-core AMD EPYC 7F52 CPU with 256 GB of RAM.
It equips with four NVIDIA A100 GPUs, each has 40 GB of on-device memory. 
The host runs on CentOS 8, and our toolchain consists of
NVIDIA driver version \hbox{460.32.03},
CUDA 11.2,
GCC 8.3.1,
and RAJA 2022.03.

We compare and validate the numerical results produced by the CS-2 to those produced by the reference implementations.

\subsection{Performance Results}

In the following, performance results are provided for FV flux computation.
We discuss various observations to provide insights on CS-2's performance characteristics.

\begin{table}[t]
  \begin{tabular}{|c|r|l|}
  \hline
  \multicolumn{1}{|c|}{Arch/lang} & \multicolumn{1}{c|}{Avg.} & \multicolumn{1}{c|}{S.D.} \\ \hline
  Dataflow/CSL                   & 0.0823                    & 0.0000014                 \\ \hline
  GPU/RAJA              & 16.8378                   & 0.0194403                 \\ \hline
  GPU/CUDA              & 14.6573                   & 0.0111278                 \\ \hline
  \end{tabular}
  \caption{Time measurement on CS-2 and A100}
  \label{tbl:time-measurements}
  \end{table}
  
\paragraph{Measurements and comparisons}

Table~\ref{tbl:time-measurements} presents wall-clock time measurements for $1000$ applications of Algorithm~\ref{algo:tpfa} for a $750 \times 994 \times 246$ mesh.
The performance of our dataflow implementation is shown in the Dataflow/CSL row,
while the reference implementation running on an NVIDIA A100 GPU is presented in the GPU/RAJA row.
In addition, the hand-crafted CUDA version running on the same GPU is shown in the GPU/CUDA row. 
We report the average kernel time and standard deviation for each implementation from multiple runs.

The performance of our RAJA-based implementation is on par with the CUDA-based version.
It uses on average 30.79 warps per streaming multiprocessor (SM) out of the theoretical 32 warps upper bound.
It achieves a $48.11\%$ occupancy out of theoretical $50\%$ occupancy.
With an arithmetic intensity of 2.11 FLOPs/Byte, it achieves 6012 GFLOPS, $76\%$ of the peak performance.
All of these show that our reference implementations have high performance characteristics and efficiently make use of the hardware.
Our dataflow implementation, comparing to the RAJA-based reference implementation, achieves a speedup of 204x.

As reported in~\cite{MauricioSC22}, when steady state is reached during the experiments, the CS-2 consumes an average \hbox{23 kW} of power. This corresponds to \hbox{13.67 GFLOP/W} in the context of the FV flux simulation. 
In comparison, the A100 runs consume a peak of \hbox{250 W} under the same workload.
The dataflow implementation achieves a 2.2x energy efficiency with respect to the reference implementation in aggregate and without considering the host or the networking equipment.

\begin{table*}[t]
\begin{tabular}{|l|l|l|r|r|r|r|}
\hline
\multicolumn{1}{|c|}{Nx} & \multicolumn{1}{c|}{Ny} & \multicolumn{1}{c|}{Nz} & \multicolumn{1}{c|}{\begin{tabular}[c]{@{}c@{}}Total Number \\ of Cells\end{tabular}} & \multicolumn{1}{c|}{\begin{tabular}[c]{@{}c@{}}Throughput \\ {[}Gcell/s{]}\end{tabular}} & \multicolumn{1}{c|}{\begin{tabular}[c]{@{}c@{}}CS-2\\ time {[}s{]}\end{tabular}} & \multicolumn{1}{c|}{\begin{tabular}[c]{@{}c@{}}A100\\ time {[}s{]}\end{tabular}} \\ \hline
200                      & 200                     & 246                     & 9,840,000                                                                             & 121.01                                                                                   & 0.0813                                                                           & 0.9040                                                                           \\ \hline
400                      & 400                     & 246                     & 39,360,000                                                                            & 481.43                                                                                   & 0.0817                                                                           & 3.2649                                                                           \\ \hline
600                      & 600                     & 246                     & 88,560,000                                                                            & 1078.79                                                                                  & 0.0821                                                                           & 7.2440                                                                           \\ \hline
750                      & 600                     & 246                     & 110,700,000                                                                           & 1347.21                                                                                  & 0.0821                                                                           & 9.6825                                                                           \\ \hline
750                      & 800                     & 246                     & 147,600,000                                                                           & 1794.01                                                                                  & 0.0822                                                                           & 13.2407                                                                          \\ \hline
750                      & 950                     & 246                     & 183,393,000                                                                           & 2227.38                                                                                  & 0.0823                                                                           & 16.8378                                                                          \\ \hline
\end{tabular}
\caption{Results of various grid sizes for scalability experiments.}
\label{tbl:weak-scaling}
\end{table*}  

\paragraph{Scaling comparison}

We conduct a weak scaling experiment, similar to the experiment as shown in~\cite{MauricioSC22}, to demonstrate the scalability of FV flux computation. 
We modify the grid dimension along the X and Y dimensions, while keeping Z-dimension a constant.
The X and Y dimensions are grown up to the maximum size of $750 \times 994$.

Table~\ref{tbl:weak-scaling} shows the mesh size of each experiment, throughout achieved on CS-2 in Gigacell/s, the wall-clock time required to compute \hbox{$1,000$} applications of Algorithm~\ref{algo:tpfa} on CS-2, as well as the kernel time of our RAJA reference implementation running on A100.
Timing reported here correspond to computations taking place on the device only.
On CS-2, time are recorded via the SDK <time> library which uses hardware timestamp counters.
On A100, we use CUDA provided runtime functions, such as \textit{cudaEventRecord}, \textit{cudaEventSynchronize}, and \textit{cudaEventElapsedTime}, to measure the elapsed time for GPU kernel runs.
The results show that the FV flux computation also scales nearly perfectly on CS-2.


\paragraph{Data communication cost}

\begin{table}[t]
  \begin{tabular}{|c|r|r|}
  \hline
                & \multicolumn{1}{c|}{Time {[}s{]}} & \multicolumn{1}{c|}{Percentage {[}\%{]}} \\ \hline
  Data Movement & 0.0199                            & 24.18                                    \\ \hline
  Computation   & 0.0624                            & 75.82                                    \\ \hline
  Total         & 0.0823                            & 100.00                                   \\ \hline
  \end{tabular}
  \caption{Time distribution on CS-2 with our largest possible mesh size.}
  \label{tbl:time-distribution}
  \end{table}

In order to gain further insight into the performance characteristics of the FV flux computation on the CS-2, we conducted an experiment in which we modified our dataflow implementation to remove all flux computations and focus solely on data communications.
We ran this modified version of the code on a mesh of size $750 \times 994 \times 246$ for $1000$ applications of Algorithm~\ref{algo:tpfa} and obtained an average data communication time of $0.0199$ seconds.
With a total device time of $0.0823$ seconds, we can deduce that all flux computations take $0.0624$ seconds.

Table~\ref{tbl:time-distribution} presents the results of our time distribution analysis.
We observed that the data communication cost accounted for $24.18\%$ of the total device time, while the remaining $75.82\%$ was dedicated to floating-point computations.
This analysis provides a more detailed understanding of the performance characteristics of the CS-2 for FV flux computation, which can be useful for optimizing future implementations of this algorithm on this platform.

\subsection{Roofline}

The Roofline~\cite{williams_roofline_2009, yang_hierarchical_roofline_2020} model provides a visual representation of a code's performance with relative to a machine's peak performance.
This model takes into account the code's arithmetic intensity, memory bandwidth, and overall performance and presents them on a single chart.
By comparing a code's performance to the platform's performance ceiling, the Roofline model can also provide valuable insights into potential optimizations for the code.

\begin{table}[t]
  \begin{tabular}{|c||c|c|c|}
  \hline
  Operation & FLOP & Mem. traffic     & Fabric traffic \\ \hline
  60 FMUL   & 1    & 2 loads, 1 store & 0              \\ \hline
  40 FSUB   & 1    & 2 loads, 1 store & 0              \\ \hline
  10 FNEG   & 1    & 1 load, 1 store  & 0              \\ \hline
  10 FADD   & 1    & 2 loads, 1 store & 0              \\ \hline
  10 FMA    & 2    & 3 loads, 1 store & 0              \\ \hline
  16 FMOV   & 0    & 1 store          & 1 load         \\ \hline
  \end{tabular}
  \caption{Instruction and memory access counts for one mesh cell on CS-2. }
  \label{tbl:wse-operations}
  \end{table}

To compute the performance characteristics for putting our dataflow implementation on the roofline chart,
we took the same roofline model described in~\cite{MauricioSC22}.
The summary of instructions at the cell level needed for the FV flux computations,
including their floating-point operation count, memory traffic, and fabric traffic, are detailed in Table~\ref{tbl:wse-operations}.
Each PE performs computations for $N_z$ cells, with each cell having 10 faces performing a TPFA computation.
Each flux computation consists of 6 FMULs, 4 FSUBs, 1 FADD, 1 FMA, and 1 FNEG, with FMA requiring two FLOPs and other operations requiring one FLOP.
Therefore, each flux requires 14 FLOPs, and each cell performs a total of 140 FLOPs.

\begin{figure}[t]
  \centering
  \includegraphics[width=0.36\textwidth]{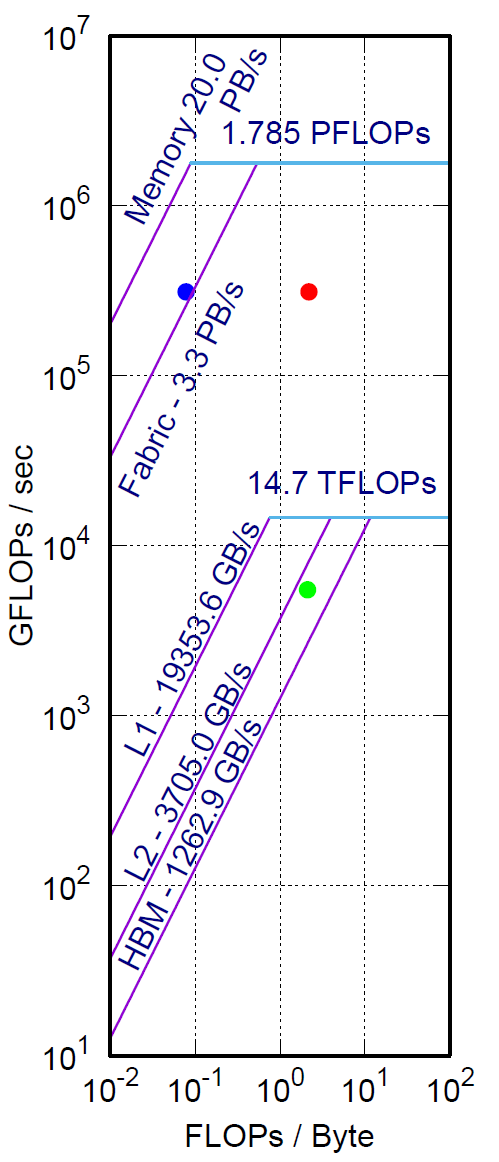}
  \caption{Roofline models for CS-2 and A100 (in log-log scale) for a $750 \times 994 \times 246$ mesh. Dots represent the implementations of FV flux computations.
 For CS-2 (top), there are two distinct resources - memory and fabric. The blue dot on the left represents memory accesses, while the red dot on the right corresponds to fabric accesses. The kernel is bandwidth-bound for memory accesses, while being compute-bound for fabric access. 
 A100 (bottom) is depicted in green. The kernel is memory-bound with achieving 76\% of the peak performance. }
  \label{figure:roofline}
\end{figure}

The floating-point operations necessitate a total of 406 loads and stores of single-precision 32-bit floating-point numbers to and from memory, and 16 loads from fabric.
These fabric loads are from the data communications of eight neighboring PEs.
Data accesses from top and bottom cells in the mesh only require memory access since they are in the same PE's memory and do not require fabric accesses.
As a result, the arithmetic intensity is 0.0862 FLOPs/Byte with respect to memory access and 2.1875 FLOPs/Byte with respect to fabric transfers.
With a mesh size of $750 \times 994 \times 246$ and the computation being completed in 0.0823s, it achieves a performance of 311.85 TFLOPs, as depicted in Figure 5 (top).
Our dataflow implementation is bandwidth-bound for memory access and compute-bound for fabric access.
However, the arithmetic intensity with respect to memory access is nearly compute-bound (0.0892 FLOPs/Byte).

For GPU, we measure our machine performance characteristics using Empirical Roofline Toolkit~\cite{yang_roofline_ert_2020} and characterize RAJA reference implementation using NVIDIA's Nsight Compute on A100. 
We present GPU roofline in Figure~\ref{figure:roofline}~(bottom).
Our RAJA implementation is memory-bound and achieve 76\% of the peak performance with respect to its arithmetic intensity.
These results also indicate that our reference implementation serves as a strong baseline for comparison purposes.

\section{Discussions}\label{sec:discussions}





This work focused on a FV flux computation using TPFA on a distributed memory system,
demonstrating the feasibility and substantial speedup achievable through this approach.
The FV flux computation is naturally extendable to a matrix-free operator FV operator for use in an iterative Krylov method which would solve equation (\ref{eq:vk}).
In combination with evaluation of sparse matrix-vector product operators, the availability of a performant matrix-free FV operator on the Cerebras architecture will be an important step in evaluating its viability for finite volume stencils.

Moreover, the algorithm presented in this study is,
to our best knowledge,
the first to exploit data communication from diagonal PEs,
which enables the implementation of other types of applications,
such as solving the acoustic wave equation on tiled transversely isotropic media,
that also require fetching data from diagonal neighbors.

Furthermore, since each PE has a distributed and isolated memory space, the memory optimization techniques discussed in this study are crucial for applications such as Reverse Time Migration workflows,
which require handling a significant amount of intermediate results.
Although the optimization process is manual, the study showcases the potential for reusing the memory space for intermediate results.





\section{Conclusions and Future Work}\label{sec:conclusion}

This study presents a novel FV flux algorithm implementation that takes advantage of a low-latency localized communications and single-level memory architecture.
This work extends localized communication patterns to exchange data from both cardinal and diagonal PEs.
We demonstrate memory-saving optimizations by reusing space for intermediate results.
We describe our strategies to speedup computations with vectorization of floating-point operations and overlapping data movements with computations.

Our experiments show that our dataflow implementation achieves 311.85 TFLOPs on a single-node CS-2 machine and
a significant speedup of 204x compared to a reference implementation running on an NVIDIA A100 GPU.
We show that the FV flux computation can as well achieve near-perfect weak scaling on a dataflow architecture.
We show additional performance and energy characteristics for the FV flux computation on a system using the dataflow architecture and demonstrate it can be exploited efficiently for this kind of applications with massive throughout and remarkable energy saving.
We also discuss the implications of this study.

Future work includes supporting arbitrary mesh topologies and mapping them efficiently onto a dataflow architecture to enable porting of a broader range of FV applications.
We also need to come up with data broadcasting strategies to support data movement from any cells in the arbitrary-shaped mesh.
Further, developing nonlinear and linear solvers on a dataflow architecture can broaden the scope of FV applications commonly used in other real-world scenarios.

\begin{acks}
The authors would like to thank TotalEnergies and Cerebras Systems for allowing to share the material.
We thank Shelton Ma and Harry Zong from TotalEnergies for all the supports.
We wish to acknowledge Leighton Wilson and Michael James from Cerebras for the introductions and instructions.

A portion of work was performed under the auspices of the U.S. Department of Energy by Lawrence Livermore 
National Laboratory under Contract DE-AC52-07NA27344.
Release LLNL-CONF-847874
\end{acks}

\pagebreak

\bibliographystyle{ACM-Reference-Format}
\bibliography{sample-base}

\appendix









\end{document}